\documentclass[12pt]{article}
\usepackage{epsfig}
\usepackage{graphics}
\usepackage{psfrag}
\usepackage{amsmath}
\usepackage{amssymb}

\newcommand{\bea}{\begin{eqnarray}}
\newcommand{\eea}{\end{eqnarray}}
\newcommand{\simgt}{\hbox{ \raise3pt\hbox to 0pt{$>$}\raise-3pt\hbox{$\sim$} }}
\newcommand{\simlt}{\hbox{ \raise3pt\hbox to 0pt{$<$}\raise-3pt\hbox{$\sim$} }}

\newcommand{\clfn}{\setcounter{footnote}{0}}

\begin{document}

\begin{titlepage}

    \begin{flushright}
      \normalsize TU-944\\
      \today
    \end{flushright}

\vskip2.5cm
\begin{center}
\Large\bf\boldmath
Estimate of 4-loop Pole-$\overline{\rm MS}$ Mass Relation\\
from Static QCD Potential
\unboldmath
\end{center}

\vspace*{0.8cm}
\begin{center}
\large
{Y. Sumino}\\[5mm]
  {\small\it Department of Physics, Tohoku University}\\[0.1cm]
  {\small\it Sendai, 980-8578 Japan}

\end{center}

\vspace*{2.8cm}
\begin{abstract}
\noindent
We estimate the presently unknown constant in the
4-loop relation between the quark pole mass and the $\overline{\rm MS}$
mass, by requiring stability of the perturbative prediction
for $E_{\rm tot}(r)=2m_{{\rm pole}}+V_{\rm QCD}(r)$
in the intermediate-distance region.
The estimate is fairly sharp due to a severe cancellation 
between $2m_{{\rm pole}}$ and $V_{\rm QCD}(r)$.
This would provide a test, 
based on general properties of
the gauge theory, for
the size of ultra-soft contributions 
to $V_{\rm QCD}(r)$.
\vspace*{0.8cm}
\noindent

\end{abstract}


\vfil
\end{titlepage}

\newpage

It has become an important theme of today's particle physics to
precisely determine the masses of heavy quarks using the frame of perturbative QCD,
as their values being indispensable inputs in various 
fields of modern particle physics.
For the purpose of precisely determining heavy quark
masses, often the relation between the pole mass and the 
mass in the modified-minimal-subtraction scheme
($\overline{\rm MS}$ mass) of a quark 
becomes necessary.
This relation can be expressed in a series expansion in
the strong coupling constant as 
\begin{eqnarray}
&& 
m_{\rm pole}
=
\overline{m}\,
\left[ 1 + d_0 \frac{ \alpha_s(\overline{m})}{\pi}
         +  d_1 \left(\frac{ \alpha_s(\overline{m})}{\pi}\right)^2
         +  d_2  \left(\frac{ \alpha_s(\overline{m})}{\pi}\right)^3
         +  d_3  \left(\frac{ \alpha_s(\overline{m})}{\pi}\right)^4
         + {\cal O} (\alpha_s^5 )
\right] \, .
\nonumber\\
\label{m-pole}
\end{eqnarray}
Here,
$\overline{m} \equiv m_{\overline{\rm MS}}(m_{\overline{\rm MS}})$
denotes the $\overline{\rm MS}$ mass renormalized at the
$\overline{\rm MS}$ mass scale;
$\alpha_{s}(\mu)=\alpha_s^{(n_l)}(\mu)$ represents
the strong coupling constant in the $\overline{\rm MS}$ scheme, 
where $n_l$ is the number of light quark flavors
($n_l=3$, 4 and 5 for the charm, bottom and top quarks, respectively);
the renormalization scale $\mu$ is set to $\overline{m}$.
For the purpose of the analysis in this paper, we use the
coupling constant of the theory with
$n_l$ flavors only as the expansion parameter.
The one-loop coefficient is given by $d_0=4/3$.
The coefficients $d_1$ and $d_2$ are obtained from the two-loop \cite{2loop-mass-relation}
and three-loop \cite{3loop-mass-relation}\footnote{
The same relation was obtained before
in \cite{Chetyrkin-Steinhauser} in a certain approximation.
} mass 
relations in the full theory (with $n_h$ heavy quarks and $n_l$ light quarks),
respectively, by rewriting them in terms of the coupling constant of the theory with
$n_l$ light quarks only.\footnote{
This relation coincides with Eq.~(14) of \cite{3loop-mass-relation}.
Note that in the other formulas of \cite{3loop-mass-relation}, the coupling 
constant of the full theory is used.
} 
At present only limited part of $d_3$ are known \cite{bb,Lee:2013sx},
and there have been increasing demands for its
full evaluation recently.

Major estimates of $d_3$ which have been performed so far rely on the
renormalon dominance hypothesis of the pole mass
\cite{bb,pinedaJHEP,Ayala:2012yw,Bali:2013pla}.
(This includes the estimate in the so-called ``large-$\beta_0$ approximation.'')
In these methods, there is an assumption (with certain grounds, see \cite{Beneke:1998ui})
on the higher-order behavior of the perturbative expansion:
\bea
d_n \sim {\rm const.}\times n! \, n^{\beta_1/(2\beta_0^2)} 
\left(\frac{\beta_0}{2}\right)^n
~~~~~{\rm for}~~~~~n\gg 1,
\label{asymptform}
\eea
where $\beta_i$ denotes the $(i+1)$-loop coefficient of the
beta function of $\alpha_s(\mu)$.
Empirically it is known that perturbative series of many observables
are approximated well by this form even at relatively low orders.
There have also been estimates of $d_3$ in another method \cite{Kataev:2010zh}.

\begin{figure}[t]\centering
\includegraphics[width=12cm]{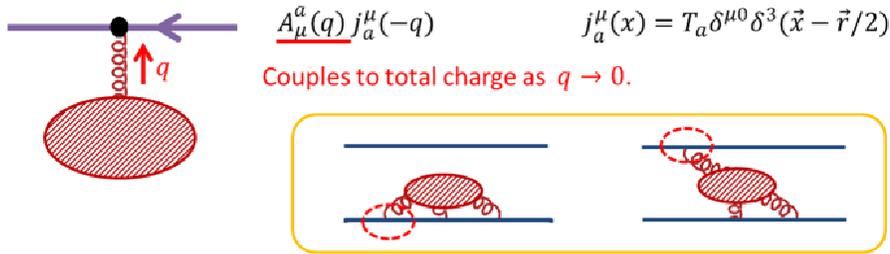}
\caption{\small
As a general feature of the gauge theory, a gluon, which couples to static 
currents $j^{\mu}_a \propto \delta^{\mu 0}$,
couples to the total charge of the system in the IR limit, $q\to 0$.
Diagrammatically both self-energy and potential-energy type diagrams
are needed for realizing this feature, hence,
for a color-singlet system, a cancellation takes place
between the two types of diagrams.
\label{FigIRgluon}
}
\end{figure}

In this paper
we present estimates of $d_3$ for $n_l=0,3,4,5$ based
on comparatively general assumptions.
In particular, our method 
does not use eq.~(\ref{asymptform}).
We consider the total energy of a color-singlet pair of heavy quarks
$Q$ and $\bar{Q}$,
defined by
\bea
E_{\rm tot}(r)=2m_{{\rm pole}}+V_{\rm QCD}(r) .
\eea
The static QCD potential
$V_{\rm QCD}(r)$ represents the potential energy between $Q$ and $\bar{Q}$
at a distance $r$, in the static limit.
We require stability of the perturbative prediction of $E_{\rm tot}(r)$
at relatively large $r$,
within the range where the perturbative prediction is
expected to be valid.
Although originally this stability was
predicted using the language of the renormalon dominance hypothesis \cite{Pineda:id},
it can be considered to hold as a general property of
perturbative QCD beyond the renormalon dominance hypothesis \cite{Beneke:1998rk}.
In fact, a gluon, which couples to static 
currents $j^{\mu}_a \propto \delta^{\mu 0}$, couples to
the total charge of the system 
$Q_a^{\rm tot}=\sum_{i} j^0_{a,i}(q=0)$ ($i=Q,\bar{Q}$)
in the zero momentum limit $q\to 0$, that is, an infra-red
(IR) gluon decouples
from the color-singlet system.
Diagrammatically an IR gluon observes the total
charge when both self-energy diagrams\footnote{
In the large mass limit contributions from IR region
to the pole mass approximate 
IR contributions to the self-energy 
of a static charge.
} 
and potential-energy diagrams
are taken into account.
Hence, a cancellation takes place between these two types of diagrams,
see Fig.~\ref{FigIRgluon}.
In perturbative QCD, convergence and stability of perturbative series
become worse as contributions from IR gluons grow.
Oppositely, after cancellation of IR contributions,
convergence and stability of perturbative predictions improve.
This can be considered as a general property of a gauge theory
which is strongly interacting at IR.

In
the perturbative series of $E_{\rm tot}(r)$ up to
${\cal O}(\alpha_s^3)$, where the exact terms
are known,
an improvement of convergence and stability
as a result of the cancellation is clearly visible,
and 
the cancellation becomes severer at higher orders.
The meaning of the latter statement is as follows.
The perturbative series of
$m_{\rm pole}$ and $V_{\rm QCD}(r)$,
respectively, do not converge well,
whereas the perturbative series of $E_{\rm tot}(r)$ 
converges much more quickly.
Let us denote the individual terms of the former as $m_n$ and $V_n$,
respectively, and of the latter as $E_n$.  
Then the ratio $|E_n/m_n|$ or $|E_n/V_n|$ reduces with $n$.
This means that there is a severer cancellation for larger $n$:
$(2 |m_n| -|V_n|)/(2 |m_n| + |V_n|) \approx E_n/(2|V_n|)$.
As a consequence, by assuming convergence and stability of $E_{\rm tot}(r)$
up to the next order
[${\cal O}(\alpha_s^4)$], we obtain fairly severe constraints on our
estimates of $d_3$.

The leading IR contributions being canceled in the static limit,
let us consider the next-to-leading IR contributions which may affect
the stability of the perturbative prediction for $E_{\rm tot}(r)$.
The interaction of the singlet static $Q\bar{Q}$ pair and
IR gluons starts from a dipole interaction $S\,\vec{r}\cdot\vec{E}^a \,O^a$
in the multipole expansion in $\vec{r}$ \cite{Brambilla:1999qa}.
The ultra-soft (US) corrections to $V_{\rm QCD}(r)$ 
originating from this interaction appear
first at ${\cal O}(\alpha_s^4)$.
It has been argued that the US corrections are small at this order \cite{Kiyo:2002rr}.
There also exist arguments that these corrections may not be small 
\cite{Ayala:2012yw}.
In our analysis, we assume that these corrections are small
and estimate $d_3$ by requiring stability of the perturbative prediction for
$E_{\rm tot}(r)$ in an IR region.\footnote{
In terms of the renormalon language, our standpoint may be phrased
as follows.
Since there exist uncanceled IR renormalons in $E_{\rm tot}(r)$,
starting from the $u=3/2$ pole in the Borel plane, they may deteriorate 
convergence of the perturbative series at higher orders of the
perturbative expansion.
Ultra-violet (UV) renormalons may also contribute.
We assume that both of these contributions are small and negligible
in estimating $d_3$.
}
Subleading IR contributions to $m_{\rm pole}$, 
which may also affect the stability of $E_{\rm tot}(r)$, 
are expected to be 
suppressed by $\Lambda_{\rm QCD}/\overline{m}$.
By increasing $\overline{m}$, we render these contributions sufficiently small.

Let us review the behavior of the perturbative series
of $E_{\rm tot}(r)$ up to ${\cal O}(\alpha_s^3)$ at relatively
large $r$, as analyzed in
\cite{Sumino:2001eh}.
Comparing the perturbative series in $\alpha_s(\mu)$ of $E_{\rm tot}(r)$
and those of
$m_{\rm pole}$ and $V_{\rm QCD}(r)$ individually,
we observe a drastic improvement in convergence of the series.
In the case $n_l=4$, $\overline{m}=4.180$~GeV and
$\alpha_s(M_Z)=0.1184$,
a stable theoretical prediction for
$E_{\rm tot}(r)$ is obtained at $r<2.8$~GeV$^{-1}$.
At each $r$, 
the scale $\mu$ is fixed in two different ways:
(1) The scale $\mu=\mu_1(r)$ is fixed by demanding stability of 
$E_{\rm tot}(r)$ against variation of the scale
(minimal-sensitivity scale \cite{Stevenson:1981vj}):
\bea
\mu \frac{d}{d\mu}E_{\rm tot}(r)\biggr|_{\mu=\mu_1(r)}=0 .
\label{scale1}
\eea
(2) The scale $\mu=\mu_2(r)$ is fixed on the minimum of the absolute value of 
the $O(\alpha_s^3)$ term $E_3$ of $E_{\rm tot}(r)$:
\bea
\mu \frac{d}{d\mu}(E_3)^2\biggr|_{\mu=\mu_2(r)}=0 .
\label{scale2}
\eea
Here and hereafter, 
we state that a stable theoretical prediction is obtained when both scales
exist;
in this case, we find that
the values of $E_{\rm tot}(r)$ corresponding to both scales agree
well, and that the convergence behaviors of both expansions
are reasonable.
The range of stable prediction extends to larger $r$ as
the order of perturbative expansion is raised [up to ${\cal O}(\alpha_s^3)$].

$E_{\rm tot}(r)$ is examined also by varying
the value of $\overline{m}$
artificially: whenever stable theoretical
predictions for $E_{\rm tot}(r)$ are obtained, the predictions
corresponding to different $\overline{m}$
agree with each other within the estimated theoretical uncertainties,
after adding an arbitrary $r$-independent constant.
[Theoretical uncertainties are estimated as order $ \Lambda_{\rm QCD}^3 r^2$ with 
$\Lambda_{\rm QCD}\simeq 300$~MeV.]
As $\overline{m}$ is increased, the perturbative predictability range of $r$, where both
scales exist, shifts to shorter-distance region.
These examinations may be regarded as tests of properties of the $SU(3)$ gauge theory,
irrespective of details of the parameters of the theory.\footnote{
In this analysis, the parameters of the theory can be taken as 
a dimensionful parameter $\Lambda_{\overline{\rm MS}}$
(which sets the unit of mass dimension)
and a dimensionless parameter $\overline{m}/\Lambda_{\overline{\rm MS}}$.
Hence, we may vary only $\overline{m}$ fixing $\Lambda_{\overline{\rm MS}}$, 
so that we can always
consider $\Lambda_{\rm QCD}$ to be of the order of 300 MeV.
}

Phenomenologically $E_{\rm tot}(r)$ is compared 
with typical phenomenological potentials.
They are in agreement in the relevant distance range, 
$0.5$~GeV$^{-1}\simlt r \simlt 2.8$~GeV$^{-1}$,
within the estimated theoretical uncertainties,
after adding an arbitrary $r$-independent constant to
each potential;
this is the case independently of the value of $\overline{m}$
(as long as a stable prediction is obtained), but
only in the cases where
realistic values are chosen for $\Lambda_{\overline{\rm MS}}$.\footnote{
These features may be summarized as follows.
Whenever a stable prediction is obtained, $E_{\rm tot}(r)$
is consistent with a function of the form 
$\Lambda_{\overline{\rm MS}}\times f(\Lambda_{\overline{\rm MS}}\, r)$ 
up to an additive constant, 
where $f(x)$ is independent of $\overline{m}$; 
only when a realistic value of $\Lambda_{\overline{\rm MS}}$ is chosen it is consistent with
phenomenological potentials.
In fact, such a function $f(\Lambda_{\overline{\rm MS}}\, r)$ 
can be explicitly extracted from $V_{\rm QCD}(r)$ 
as a short-distance dominant
(renormalon-free) part, given as
a ``Coulomb+linear'' potential \cite{Sumino:2003yp,Sumino:2005cq}.
}
There are also similar comparisons with lattice computations of $V_{\rm QCD}(r)$,
with good agreements \cite{Pineda:2002se,Brambilla:2010pp}.

\begin{figure}[t]\centering
\includegraphics[width=13cm]{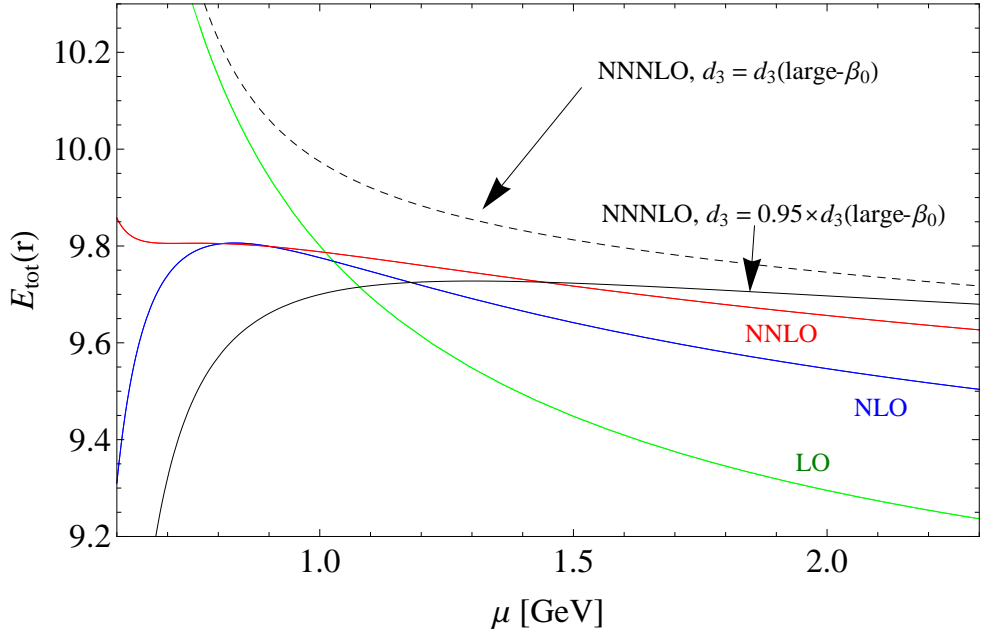}
\caption{\small
$E_{\rm tot}(r)$ at
$r=2.8$~GeV$^{-1}$ as a function of the scale $\mu$.
The solid lines represent the sum of the
perturbative series up to 
${\cal O}(\alpha_s)$ [LO], ${\cal O}(\alpha_s^2)$ [NLO],
${\cal O}(\alpha_s^3)$ [NNLO] and ${\cal O}(\alpha_s^4)$ [NNNLO,
$d_3=0.95\times d_3(\mbox{large-}\beta_0)$].
The dashed line represents the
NNNLO prediction corresponding to 
$d_3=d_3(\mbox{large-}\beta_0)$.
We set 
$\alpha_s(M_Z)=0.1184$, $\overline{m}=4.180$~GeV and $n_l=4$.
\label{Etot}
}
\end{figure}
Now we repeat the same analysis including the terms at the next order 
and varying $d_3$ in addition.
We first
set 
$\alpha_s(M_Z)=0.1184$, $\overline{m}=4.180$~GeV and $n_l=4$.
(We neglect the masses of the light quarks in the following analysis).
We take for $V_{\rm QCD}(r)$ the sum of the perturbative series up to 
${\cal O}(\alpha_s^4)$ \cite{Peter:1996ig,Anzai:2009tm} and 
${\cal O}(\alpha_s^4\log\alpha_s)$ \cite{Brambilla:1999qa,Kniehl:1999ud},
as given by eq.~(21) of \cite{Anzai:2010td};
the ${\cal O}(\alpha_s^4\log\alpha_s)$ term is generated by contributions from the US scale.
Roughly speaking, if we
choose a value of $d_3$ close to that of
the renormalon estimate \cite{pinedaJHEP} or to the large-$\beta_0$ value \cite{bb}
\bea
d_3(\mbox{large-}\beta_0)  &\simeq & 
3046.29 - 553.872\,{n_l} +   33.568\,{{{n_l^2}}} - 0.678141\,{{{n_l^3}}},
\label{d3largebeta0}
\eea
a cancellation between 
$2m_{\rm pole}$ and $V_{\rm QCD}(r)$ takes place and
a relatively convergent and stable prediction is obtained.
Nevertheless, the level of cancellation depends sensitively
on the value of $d_3$.
For demonstration we show in Fig.~\ref{Etot} 
the scale dependences of $E_{\rm tot}(r)$ at
$r=2.8$~GeV$^{-1}$, 
and $d_3=0.95\times d_3(\mbox{large-}\beta_0)$.
Four solid lines are plotted, corresponding to the sum of the
perturbative series up to 
${\cal O}(\alpha_s)$ [LO], ${\cal O}(\alpha_s^2)$ [NLO],
${\cal O}(\alpha_s^3)$ [NNLO] and ${\cal O}(\alpha_s^4)$ [NNNLO].
The next-to-next-to-next-to-leading order
(NNNLO) prediction corresponding to 
$d_3=d_3(\mbox{large-}\beta_0)$ is also shown with a dashed line.

As already stated, at NNLO both scales $\mu_1(r)$ and $\mu_2(r)$
[eqs.~(\ref{scale1})
and (\ref{scale2})] exist up to $r\simeq 2.8$~GeV$^{-1}$.
We require that at NNNLO both scales also exist at least
up to the same $r$, such that the perturbative stability
is not deteriorated at this order.
This requirement leads to an upper bound for $d_3$:
$d_3< 0.96\times d_3(\mbox{large-}\beta_0)$.
We also vary $\overline{m}$ artificially to 8~GeV and 16~GeV
and require that at NNNLO the two scales exist at least
up to the same $r$ as at NNLO.  
(The corresponding values of $r$ are
1.4~GeV$^{-1}$ and 0.7~GeV$^{-1}$, respectively.)
In these cases, a common value 
$0.97\times d_3(\mbox{large-}\beta_0)$
is obtained as upper bounds for $d_3$.
All the upper bounds are fairly solid, in the sense that as soon
as we assign a larger value to $d_3$ in each case, 
we observe a strong instability
of the perturbative prediction; see Fig.~\ref{Etot}.
Since all the upper bounds are of similar values, 
we consider that effects of $1/\overline{m}$-suppressed
contributions to $m_{\rm pole}$ are sufficiently small and take 
$0.97\times d_3(\mbox{large-}\beta_0)$ as the reference value for the
upper bound of $d_3$ of our estimate.

\clfn

\begin{table}[t]
\begin{center}
\begin{tabular}{c|cccccl}
\hline
$n_l$ &\small
Ref.~\cite{bb} & \small Ref.~\cite{pinedaJHEP} &\small Ref.~\cite{Ayala:2012yw} 
&\small Ref.~\cite{Kataev:2010zh}&\small Ref.~\cite{Bali:2013pla}
&\small
 \hspace*{18mm}Our estimate\\
\hline
0 & 3046.29 & 3706.78 & -- & -- & 3933.01 
& \!\!$(1.1\pm 0.05)\!\times\! d_3(\mbox{large-}\beta_0)
\approx 3351^{+152}_{-152}$ \\
3 & 1668.48 & 1818.60 & 1785.9 & 1281 & -- & $(1.0\pm 0.1)\!\times\! d_3(\mbox{large-}\beta_0)
\approx 1668^{+167}_{-167}$ \\
4 & 1324.49 & 1345.72 & 1316.4 & ~986 & -- & $(0.95^{+ 0.02}_{-0.05})\!\times\! d_3(\mbox{large-}\beta_0)
\approx \, 1258^{~+26}_{~-66}$ \\
5 & 1031.37 & ~947.90 & ~920.1 & ~719 & -- & $(0.87^{+ 0.03}_{-0.17})\!\times\! d_3(\mbox{large-}\beta_0)
\approx ~~897^{~+31}_{-175}$ \\
\hline
\end{tabular}
\caption{\small
Comparison of different estimates of $d_3$ defined in eq.~(\ref{m-pole}).
The estimate of Ref.~\cite{bb} denotes $d_3(\mbox{large-}\beta_0)$; 
those of Refs.~\cite{pinedaJHEP},
\cite{Ayala:2012yw} and \cite{Bali:2013pla}
are based on the renormalon hypothesis;
the estimate of Ref.~\cite{Kataev:2010zh} is derived by an effective charge method.
}
\label{results}
\end{center}
\vspace*{-3mm}
\end{table} 
If we assign a value much smaller than $d_3(\mbox{large-}\beta_0)$ to $d_3$, 
qualitatively the perturbative series of $E_{\rm tot}(r)$ at NNNLO 
tends to become
unstable and exhibit a poorer convergence behavior.
For instance, the scales  fixed 
at NNLO and NNNLO [eq.~(\ref{scale1}) or eq.~(\ref{scale2})]
tend to be separated farther;
the crossing points of the solid lines in Fig.~\ref{Etot},
which are centered to a small region, tend to
be separated apart.
We quantify this feature by further demanding
that the difference between the NNLO and NNNLO predictions 
be smaller than the perturbative uncertainty\footnote{
The estimate of order $\Lambda_{\rm QCD}^3\,r^2$ is among the predictions of the renormalon 
dominance hypothesis.
It can, however, be derived also in a more general framework, without assuming
the renormalon dominance, eq.~(\ref{asymptform}).
Namely, within the effective field theory ``potential non-relativistic QCD''
(pNRQCD), the leading IR contribution to $E_{\rm tot}(r)$ can be 
absorbed into a matrix element of a non-local gluon condensate, whose size is of
order $\Lambda_{\rm QCD}^3\,r^2$ by dimensional analysis \cite{Brambilla:1999qa}.
In this case only UV contributions remain in the
perturbative expansion of the Wilson coefficient, which should be 
more convergent than the perturbative series of $E_{\rm tot}(r)$.
Thus, IR contributions to $E_{\rm tot}(r)$ generates (at most) order 
$\Lambda_{\rm QCD}^3\,r^2$
uncertainties.
(See also \cite{Sumino:2005cq}.)
}
$\Lambda_{\rm QCD}^3\,r^2$ ($\Lambda_{\rm QCD}=300$~MeV)
for each of the scale choices $\mu=\mu_1(r)$ and $\mu=\mu_2(r)$;
since the estimate $\Lambda_{\rm QCD}^3\,r^2$ is  meaningful only in an IR region,
we apply this requirement in the range $r>1$~GeV$^{-1}$.
This requirement sets a lower bound for $d_3$
corresponding to each value of $\overline{m}$.\footnote{
Since we require consistency with the NNLO prediction, the perturbative
prediction of $E_{\rm tot}(r)$ at NNNLO
also agrees with typical phenomenological potentials if we take a realistic value
for $\Lambda_{\overline{\rm MS}}$.
}
We obtain $d_3\simgt 0.90\times d_3(\mbox{large-}\beta_0)$. 
This value, however, depends on our choice 
$\Lambda_{\rm QCD}=300$~MeV.
Thus, in comparison to the upper bound, the lower bound is
to some extent obscure.

After choosing a value of $d_3$ within the range determined 
by the above two requirements, qualitatively
the perturbative prediction for
$E_{\rm tot}(r)$ becomes stable and the perturbative series
exhibits an optimally convergent
behavior.
This effect is enhanced especially at larger $r$.
An optimal estimate is 
$d_3\approx 0.95\times d_3(\mbox{large-}\beta_0)$; see Fig.~\ref{Etot}.
We repeat the same analyses for $n_l=0$, 3 and 5 and find
qualitatively similar results.
Our estimates of $d_3$ are summarized in Table~\ref{results}.
Other estimates of $d_3$ are also listed in the same table for comparison.
We note that the large-$\beta_0$ values and some of the renormalon estimates for
$n_l=4,5$
lie above the upper bounds of our estimates.

Once $d_3$ is computed exactly in the future, a comparison with our estimates will
test our understanding of the perturbative series of $E_{\rm tot}(r)$.
In particular, it will test our
assumption that the US contributions (the
leading residual IR contributions in the multipole expansion 
that may affect the stability
of the perturbative series) are small and
do not deteriorate the perturbative convergence observed up to NNLO.
Other assumptions are based on general properties of a
gauge theory strongly interacting at IR and are independent 
of the renormalon dominance eq.~(\ref{asymptform}).
We obtained 
fairly constrained estimates for $d_3$
reflecting a severe cancellation between $2m_{\rm pole}$ and
$V_{\rm QCD}(r)$.

\section*{Acknowledgements}
The author is grateful to A.~Pineda and
Y.~Kiyo for fruitful
discussion.
This work is supported in part by 
Grant-in-Aid for
scientific research (No.\ 23540281) from
MEXT, Japan.


\end{document}